\documentclass[a4paper,10pt,twoside,reqno,nonamelimits]{article}
\usepackage{a4wide}
\usepackage[colorlinks=true,urlcolor=blue,linkcolor=blue,citecolor=blue]{hyperref}
\usepackage{newcent}         

\usepackage{amsmath}
\usepackage{amssymb}
\usepackage{amsfonts}
\usepackage{amscd}
\usepackage{amsthm}
\usepackage{amsxtra}
\usepackage{mathrsfs}
\usepackage{nicefrac}
\usepackage{graphicx}
\usepackage{xcolor}
\usepackage{qcircuit}
\usepackage{tikz}
\usepackage{enumitem}
\setlist[enumerate]{leftmargin=2em,itemindent=0em, labelindent=0pt,labelwidth=1.5em,labelsep=.5em, align=left, noitemsep}
\newlist{txtenum}{enumerate}{1}
\setlist[txtenum]{leftmargin=0em,itemindent=1.5em, labelindent=0pt,labelwidth=1em,labelsep=.5em, align=left}

\usepackage{algorithm2e}\DontPrintSemicolon
\SetKwFor{QCircuit}{Circuit}

\theoremstyle{plain}
\newtheorem{theorem}{Theorem}
\newtheorem*{theorem*}{Theorem}

\newtheorem{proposition}[theorem]{Proposition}
\newtheorem*{proposition*}{Proposition}

\newtheorem*{corollary*}{Corollary}

\newtheorem*{lemma*}{Lemma}

\newtheorem*{observation*}{Observation}

\newtheorem*{conjecture*}{Conjecture}

\newtheorem*{question*}{Question}

\newtheorem*{questions*}{Questions}

\newtheorem*{problem*}{Problem}

\newtheorem*{problems*}{Problems}

\newtheorem*{openproblem*}{Open Problem}

\theoremstyle{definition}

\newtheorem*{definition*}{Definition}

\newtheorem*{example*}{Example}

\newtheorem*{exercise*}{Exercise}

\newtheorem{remark}[theorem]{Remark}
\newtheorem*{remark*}{Remark}

\newtheorem*{remarks*}{Remarks}

\theoremstyle{remark}

\newtheorem*{claim*}{Claim}

\newcommand{\subclass}[1]{}

\newcommand{\enumTi}[1]{\renewcommand{\theenumi}{#1}}

\newcommand{\alphenumi}{\enumTi{\alph{enumi}}}

\newcommand{\romenumi}{\enumTi{\roman{enumi}}}

\newlength{\hspaceforlengthglumpf}

\newcommand{\comment}[1]{\text{\footnotesize[#1]}}

\newcommand{\Id}{\mathrm{Id}}

\newcommand{\One}{\mathbf{1}}

\newcommand{\lt}{\left}
\newcommand{\rt}{\right}

\newcommand{\abs}[1]{{\lt\lvert{#1}\rt\rvert}}

\newcommand{\nfrac}[2]{{\nicefrac{#1}{#2}}}

\newcommand{\bra}[1]{{\lt< #1 \rt|}}
\newcommand{\ket}[1]{{\lt| #1 \rt>}}

\newlength{\algotabbingwidth}
\begin{document}
\title{Note on (active-)QRAM-style data access as a quantum circuit}
\author{Tore Vincent Carstens and Dirk Oliver Theis\thanks{Partly supported by the Estonian Research Council, ETAG (\textit{Eesti Teadusagentuur}), through PUT Exploratory Grant \#620.}\\[1ex]
  \small Institute of Computer Science, University of Tartu\\
  \small {\tiny and} Ketita Labs {\tiny O\"U}\\
  \small Tartu, Estonia\\
  \small \{\texttt{carstens}, \texttt{dotheis}\}\texttt{@}\{\texttt{ketita.com}, \texttt{ut.ee}\}%
}
\date{May 2018}
\maketitle

\begin{abstract}
  We observe how an active (i.e., requring $2^n$ parallel control operations) QRAM-like effect
  $$
  \sum_{y=0}^{N-1} \ket{y}\bra{y} \otimes U^y_{\text{result},\text{memory}_y}
  $$
  can be realized, as a quantum circuit of depth $O(n+\sqrt m)$ (where $m$ is the size of the result register) plus the maximum over all~$z$ of the circuit depths of controlled-$U^z$ operations.
  \par\medskip%
  \textbf{Keywords:} Gate-based quantum computing, quantum circuits, QRAM.
\end{abstract}

\newcommand{\emptilon}{\varepsilon}
\newcommand{\address}{\texttt{address}}
\newcommand{\result}{\texttt{result}}
\newcommand{\res}{\texttt{res}}
\newcommand{\life}{\texttt{life}}
\newcommand{\addr}{\texttt{adr}}
\newcommand{\mem}{\texttt{mem}}
\newcommand{\verboseProof}[1]{}                           

\section{Introduction}\label{sec:intro}
All data processing, classical or quantum, requires that input data be made available in the computer.
On a quantum computer, one way to make data available is through Quantum Random Access Memory (QRAM).
QRAM is an abstract concept defined through the following interaction: If $y$ is a nonnegative $n$-bit integer representing the \textit{address} of a memory location, $\ket{\mu_y} = \sum_{s\in\{0,1\}^m}\alpha_{s_y} \ket{s}$ is a pure $m$-qubit quantum state ``stored'' at that ``memory location'', and $r\in\{0,1\}^m$ is a target $m$-bit string, then QRAM access has the following effect:
\begin{equation*}
  \ket{y}\ket{r}\ket{\mu_y} \xrightarrow{\text{QRAM access}} \ket{y} \sum_{s\in\{0,1\}^m} \alpha_{s_y} \ket{s\oplus r} \ket{s},
\end{equation*}
where $\oplus$ is the bit-wise XOR between bit-strings.
This can be concisely written as
\begin{equation*}
  \sum_{y=0}^{N-1} \ket{y}\bra{y} \otimes CNOT^{\otimes m}_{\mem_y,\result}.
\end{equation*}

QRAM is an implicit assumption that quantum algorithms such as the quantum linear system solver, HHL~\cite{harrow-hassidim-lloyd:hhl:2009}, and quantum machine learning algorithms derived from it make on the quantum computer on which they run.  A physical realization of QRAM with $O(n^2)$ access time has been proposed \cite{Giovannetti-Lloyd-Maccone:QRAM:2008,Giovannetti-Lloyd-Maccone:QRAM-arch:2008}; it is, however, unclear whether it is possible to build a so-called ``passive'' QRAM~\cite{aaronson:fine-print:2015}, i.e., one which doesn't require $2^n$ parallel (classical) operations controling the quantum hardware.  Using active QRAM will not give exponential speedup of, say, HHL over classical algorithms using $2^n$ processors.

This paper deals with a realization of (an obvious generalization of) active QRAM as a quantum circuit.  Given, for each $z\in\{0,\dots,2^n-1\}$, a ``memory register'' $\mem_z$ of~$k_z$ qubits and a unitary operation $U^z$ acting on two registers, a \textit{result register} of~$m$ qubits and the said ``memory register'', we realize the unitary operation defined through
\begin{equation}
  \ket{y}\ket{\result}\ket{\mem_y} \mapsto \ket{y} U^y(\ket{\result}\ket{\mem_y}),
\end{equation}
whenever $y\in\{0,\dots,2^n-1\}$ and $\ket{\result}$ is in a computational basis state.
This can be written concisely as an operation on the Hilbert space $\displaystyle \mathcal H^{(n)}_{\address} \otimes H^{(m)}_{\result} \otimes \bigotimes_{z=0}^{N-1} H^{(k_z)}_{\mem_z}$ (the superscripts give the number of qubits):
\begin{equation}\label{eq:fundamental-op}
  \sum_{y=0}^{N-1} \ket{y}\bra{y} \otimes U^y
\end{equation}
where it is understood, by abuse of notation, that $U^y$ acts on $H_{\result} \otimes H_{\mem_y}$, i.e., it is tensored with identities on $H_{\mem_z}$, $z\ne y$.

The point of this note is the observation that at the expense of (a) $2^n$ parallel quantum operations being performed, and (b) $O(m 2^n)$ ancillary qubits, \eqref{eq:fundamental-op} can be realized in with a quantum circuit of depth $n + \sqrt m$ plus the maximum (over all~$z$) circuit depth of a controlled-$U^z$.

\medskip%
Taking $k_z=m$ for all~$z$, and $U^z$ an $m$-fold tensor-product of CNOTs gives QRAM.  Given a function $f\colon \{0,1\}^n \to \{0,1\}^m$, setting, $k_z=0$ for all~$z$, and
\begin{equation*}
  U^z := \bigotimes_{i=0}^m X^{f(z)_i}
\end{equation*}
(where $X$ stands for the Pauli $X$ operator and exponents are taken as usual) realizes the unitary $U_f$ with $U_f(\ket{y}\ket{r}) = \ket{y}\ket{f(y)\oplus r}$.

Hence, it can be said that \eqref{eq:fundamental-op} gives access to data which is partly ``hard-coded'' into the quantum circuit, and partly ``stored'' in qubits on the quantum processor.  Another example are, e.g., controlled rotations $e^{-i\pi \mu X}$, where $\mu$ is an $m$-bit fixed point fraction stored in $m$~qubits (possibly in superposition).

\medskip%
Our proposed quantum circuit follows the structure of \cite{Giovannetti-Lloyd-Maccone:QRAM:2008}, i.e., it is arranged in a binary tree in such a way that operations on nodes with the same distance from the root can be run in parallel.

\section{Description of the quantum circuit}
\subsubsection*{Some notation}
Let $N := 2^n$.
We freely switch between interpreting nonnegative integers in $\{0,\dots,N-1\}$ as bit-strings of length~$n$; as usual, bit-strings have the higher-significant bits to the left.

The empty bit-string is denoted by $\emptilon$.

\subsubsection*{Overview}
Suppose implementations (quantum circuits or ``black boxes'') of the controlled unitaries
\begin{equation}\label{eq:ctrl_Uz}
  \ket{0}\bra{0}\otimes \Id + \ket{1}\bra{1}\otimes U^z
\end{equation}
for $z \in\{0,1\}^n$ are given.  Each $U^z$ acts on two quantum registers: a result register $\res_z$ of size $m$, and a ``memory'' register $\mem_z$ of size $k_z$; we allow $k_z=0$.  All these registers $\res_z$, $\mem_z$, $z\in\{0,1\}^n$ are assumed disjoint.

As in \cite{Giovannetti-Lloyd-Maccone:QRAM:2008}, the whole process is organized in a binary tree.  The nodes of the tree are labeled by bit-strings $x = x_{\ell-1}\cdots x_0$ of up to $n$ bits; we denote by $\abs{x}$ length of the bit-string ($\ell$ in the case of $x = x_{\ell-1}\cdots x_0$).  The root of the tree has the label $\emptilon$, which is the empty bit-string; if $x$ labels a node and $\abs{x}<n$, then $x0$ and $x1$ are the labels of the two (left and right) children of~$x$; the leaves of the tree are the bit- strings of length~$n$.

\subsection{Down--Run--Up}
As in \cite{Giovannetti-Lloyd-Maccone:QRAM:2008}, the quantum circuit operates in two phases: The ``Down'' phase, which propagages the address information from the root of the tree to the leaves; and the ``Up'' phase, which propagates the result of running $U^z$ back to the root.  Between the two, we have a ``Run'' phase, which runs the controlled unitaries~\eqref{eq:ctrl_Uz}.

Uncomputation is needed in general, i.e., the complete quantum circuit will be: Down--Run--Up--do-stuff--(Down--Run--Up)$^\dagger$.

\subsubsection{The ``Down'' phase}
For each non-root node~$x$ in the tree (i.e., each bit-string $x$ with $1\le \abs{x}\le n$), we use an ancilla qubit $\life_x$.  If the address register $\address$ is in a computational basis state $\ket{z}$, then the ``Down'' phase will set $\life_x$ to state $\ket1$, iff the node~$x$ is on the path from the root to the leaf with label~$z$.

For each non-leaf node $x$ in the tree (i.e., each bit-string $x$ with $1\le \abs{x} < n$), we use an ancilla register $\addr_x$ with $n-\abs{x}$ qubits.  If the address register $\address$ is in a computational basis state $\ket z$ with $z=z_{n-1}\cdots z_0$, then the ``Down'' phase will set $\addr_x$ to the state $\ket{z_{n-\abs{x}-1}\cdots z_0}$, i.e., $\addr_x$ is a copy of the $n-\abs{x}$ least significant bits of the address register.

We also need to ``hand down'' the contents of the result register $\res$.  For that, for each non-root, non-leaf tree node~$x$, we use an ancilla register $\res_x$, of size~$m$.  This is in addition to the register $\res_z$, for each leaf~$z$, on which the $U^z$ operate, which we also consider as ancilla registers.  Further, we denote $\res_\emptilon := \result$.

The ``Down'' phase proceeds as follows.  First of all, all ancilla qubits (including $\res_z$, for $z$ of length~$n$) are prepared in state $\ket0$, except for $\life_\emptilon$, whis is prepared in state $\ket 1$.

\begin{center}
\fbox{%
\begin{minipage}{.9\linewidth}
For each $k=0,2,3,\dots,n-1$ (sequentially) do the following: \begin{enumerate}[label=\arabic*.)]
\item For each node with label $x$ of length $k$ in parallel:\\                      
  for each $j=0,\dots,n-k-1$ in parallel:\\
  Apply the following CNOT gate:
  Controlled on $\addr_x[j]$ flip $\addr_{x0}[j]$
\item For each node with label $x$ of length $k$ in parallel:\\                      
  for each $j=0,\dots,n-k-1$ in parallel:\\
  Apply the following CNOT gate:
  Controlled on $\addr_x[j]$ flip $\addr_{x1}[j]$
\item For each node with label $x$ of length $k$ in parallel:\\                      
  Apply the following Toffoli gate:\\
  Controlled on $\addr_x[n-k-1]$ and on $\life_x$, flip $\life_{x0}$.
\item For each node with label $x$ of length $k$ in parallel:\\                      
  Sandwitched between two applications of the Pauli-X gate on $\addr_x[n-k-1]$, apply the following Toffoli gate:\\
  Controlled on $\addr_x[n-k-1]$ and on $\life_x$, flip $\life_{x1}$.
\item\label{step:res-swap}
  For each node with label $x$ of length $k$ in parallel:\\                          
  for each $i=1,\dots,m$ sequentially:\\
  Apply the following two Fredkin gates in parallel:\\
  Controlled on $\life_{x0}$, swap $\res_{x}[i]$ and $\res_{x0}[i]$; and\\
  Controlled on $\life_{x1}$, swap $\res_{x}[i]$ and $\res_{x1}[i]$.
\end{enumerate}
\end{minipage}%
}%
\end{center}

This has the effect that the address bits required to determine the \life{}ness of each node is ``handed down'' in the tree to the leaves, which allows, on each level~$k$ of the tree, to run in parallel the operations for all nodes on that level.

At the end of the ``Down'' phase, $\life_z$, $z\in \{0,1\}^n$, indicates whether the execution of $U^z$ is requested.

\subsubsection{Resource analysis}
The \life-qubits alone already require $~2N$ qubits.  A short calculation shows that the number of $\addr$-qubits is
\begin{equation*}
  \sum_{k=0}^{n-1} (n-k-1) 2^k = (1+o(1))\, N
\end{equation*}
\verboseProof{
  Indeed, we have
  \begin{align*}
    \sum_{k=0}^{n-1} (n-k-1)2^k
    &= 2^{n-1}   \sum_{\ell=0}^{n} \ell z^\ell     && \comment{$\ell:=n-k-1$, $z := \nfrac12$}\\
    &= 2^{n-1} z \sum_{\ell=0}^{n} \ell z^{\ell-1} && \\
    &= 2^{n-2}   \partial_z \frac{1-z^n}{1-z}      && \\
    &= 2^{n-2}   \frac{-nz^{n-1}(1-z) + (1-z^n)}{(1-z)^2} && \\
    &= 2^{n-2}   \frac{1 + z^n\,((1-\nfrac1z)n - 1)}{(1-z)^2} && \\
    &= 2^{n-2} \frac{1+o(1)}{1/4}                            && \\
    &=(1+o(1))\, N.
  \end{align*}
}
The number of $\res$-qubits is, of course, $(1-o(1))2mN$.  In total, the number of ancilla qubits is $(1+o(1))(2m+3)N$.

Owing to the parallelism, the circuit depth is $O(n+m)$.  The circuit width (maximum number of parallel operations) is $2mn$.

\begin{remark}
  At the expense of $\sqrt{m}$ additional ancilla qubits, the circuit depth can be reduced to $O(\sqrt{m} + n)$.
  Indeed, replace the sequential loop in step\ref{step:res-swap} by the following.  In every step $i = 1,\dots,\sqrt m$, do this in parallel: (a) make a CNOT-copy of the control-ancilla, and (b) perform $s-1$ controlled operations controlled on the copies of the ancilla qubit created in the earlier steps.  Finally, uncompute the ancillas.
\end{remark}

\subsubsection{The ``Run'' phase}
After the completion of the ``Down'' phase, we execute the controlled unitaries~\eqref{eq:ctrl_Uz}.

For each $z\in\{0,1\}^n$ in parallel:
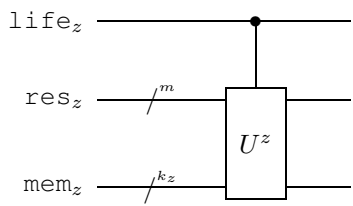
\begin{equation*}
  \Qcircuit{
    \lstick{\life_z} & \qw                & \ctrl{1}           & \qw \\
    \lstick{\res_z}  & {/}^{{}^m}   \qw   & \multigate{1}{U^z} & \qw \\
    \lstick{\mem_z}  & {/}^{{}^{k_z}} \qw &        \ghost{U^z} & \qw \\
  }
\end{equation*}
Recall that~$k_z$ can be zero.

\subsubsection{The ``Up'' phase}
The ``Up'' phase moves the result from the leaves up to the root.  As a unitary operator, it is, the adjoint operation of the ``Down'' phase.  We repeat it here for clarity.

\begin{center}
\fbox{%
\begin{minipage}{.9\linewidth}
For each $k=n-1,n-2,\dots,0$ (sequentially) do the following:
\begin{enumerate}[label=\arabic*.)]
\item For each node with label $x$ of length $k$ in parallel:\\
  for each $i=1,\dots,m$ in parallel:\\
  Apply the following two Fredkin gates in parallel:\\
  Controlled on $\life_{x0}$, swap $\res_{x}[i]$ and $\res_{x0}[i]$; and\\
  Controlled on $\life_{x1}$, swap $\res_{x}[i]$ and $\res_{x1}[i]$.

\item For each node with label $x$ of length $k$ in parallel:\\                      
  Sandwitched between two applications of the Pauli-X gate on $\addr_x[n-k-1]$, apply the following Toffoli gate:\\
  Controlled on $\addr_x[n-k-1]$ and on $\life_x$, flip $\life_{x1}$.

\item For each node with label $x$ of length $k$ in parallel:\\                      
  Apply the following Toffoli gate:\\
  Controlled on $\addr_x[n-k-1]$ and on $\life_x$, flip $\life_{x0}$.

\item For each node with label $x$ of length $k$ in parallel:\\                      
  for each $j=0,\dots,n-k-1$ in parallel:\\
  Apply the following CNOT gate:
  Controlled on $\addr_x[j]$ flip $\addr_{x1}[j]$

\item For each node with label $x$ of length $k$ in parallel:\\                      
  for each $j=0,\dots,n-k-1$ in parallel:\\
  Apply the following CNOT gate:
  Controlled on $\addr_x[j]$ flip $\addr_{x0}[j]$

\end{enumerate}
\end{minipage}%
}%
\end{center}

\subsection{The effect}
It should be apparent from the construction that the circuit does what it is supposed to do:
\begin{proposition}
  If the address register is in a computational basis state $\ket{\address}=\ket{y}$, for $y\in\{0,1\}^n$, then the unitary transformation has the following effect:
  \begin{equation}
    \ket{\address} \ket{\result, \mem} \underbrace{\ket{0}\cdots\ket{0}}_{\text{ancillas}}
    \longrightarrow
    \ket{\address} (U^y\otimes \One)(\ket{\result, \mem}) \ket{0}\cdots\ket{0},
  \end{equation}
  where the effect of $(U^y\otimes \One)$ is $U^y$ on $\result$ and $\mem_y$ registers, and identity on all $\mem_z$ with $z\ne y$.
\end{proposition}


\bibliographystyle{plain}
\bibliography{dirks}
\end{document}